\documentclass[10.5pt,a4paper]{article}
\usepackage[dvipdfmx]{graphicx}
\usepackage{subfigure}
\usepackage{here}
\graphicspath{{picture/}} 

\usepackage{enumerate}
\usepackage{paralist}

\makeatletter
\def\Hline{\noalign{\ifnum0=`}\fi\hrule\@height 3.\arrayrulewidth \futurelet\reserved@a\@xhline}
\makeatother

\usepackage[top=30truemm,bottom=30truemm,left=20truemm,right=20truemm]{geometry}
\usepackage{fancyhdr}
\usepackage{amsmath}
\usepackage{amssymb}
\usepackage{mathcomp}
\usepackage{siunitx} 
\usepackage{cases}

\usepackage{authblk} 
\usepackage{color}

\usepackage{comment}

\title{Improvement of the target sensitivity in DECIGO by optimizing its parameters for quantum noise including the effect of diffraction loss}

\author[1,$\dagger$]{Tomohiro Ishikawa}
\author[1]{Shoki Iwaguchi}
\author[2]{Yuta Michimura}
\author[2]{Masaki Ando}
\author[1]{Rika Yamada}
\author[1]{Izumi Watanabe}
\author[3]{Koji Nagano}
\author[4]{Tomotada Akutsu}
\author[3]{Kentaro Komori}
\author[5]{Mitsuru Musha}
\author[1]{Takeo Naito}
\author[1]{Taigen Morimoto}
\author[1]{Seiji Kawamura}

\affil[1]{Department of Physics, Nagoya University, Nagoya, Aichi 464-8602, Japan}
\affil[2]{Department of Physics, The University of Tokyo, Bunkyo, Tokyo 113-0033, Japan}
\affil[3]{Institute of Space and Astronautical Science, Japan Aerospace Exploration Agency, Sagamihara, Kanagawa 252-5210, Japan}
\affil[4]{Gravitational Wave Science Project, National Astronomical Observatory of Japan, Mitaka, Tokyo 181-8588, Japan}
\affil[5]{Institute for Laser Science, The University of Electro-Communications, Chofu, Tokyo 182-8585, Japan}

\date{}

\begin{document}
\maketitle

\begin{abstract}
\noindent
  DECIGO is the future Japanese gravitational wave detector in outer space. We previously set the default design parameters to provide a good target sensitivity to detect the primordial gravitational waves (GWs). However, the updated upper limit of the primordial GWs by the Planck observations motivated us for further optimization of the target sensitivity. Previously, we had not considered optical diffraction loss due to the very long cavity length. In this paper, we optimize various DECIGO parameters by maximizing the signal-to-noise ratio (SNR), for the primordial GWs to quantum noise including the effects of  diffraction loss. We evaluated the power spectrum density for one cluster in DECIGO utilizing the quantum noise of one differential Fabry-Perot interferometer. Then we calculated the SNR by correlating two clusters in the same position. We performed the optimization for two cases: the constant mirror-thickness case and the constant mirror-mass case. As a result, we obtained the SNR dependence on the mirror radius, which also determines various DECIGO parameters. This result is the first step toward optimizing the DECIGO design by considering the practical constraints on the mirror dimension and implementing other noise sources.\\

\end{abstract}

\section{Introduction}
The existence of gravitational waves (GWs) was predicted by Einstein's theory of general relativity  and verified recently by LIGO and Virgo \cite{LIGO} \cite{LIGOVirgo}. GWs are propagating spacetime waves produced by changes in the distribution of mass/energy. Examples of GW sources include the merger of black hole binaries and that of neutron star binaries \cite{BHBH}. Among various origins of GWs, inflation in the early universe could have produced a stochastic background of primordial GWs through the quantum fluctuations in spacetime \cite{PGWs}. The detection of the primordial GWs is expected to reveal how our universe began. However, it is very challenging to observe the primordial GWs with ground-based detectors, because the waves' magnitude is too small in the ground-based detector's frequency band (\SI{10}{Hz} - \SI{10}{kHz}) \cite{PGWs}.

\begin{figure}[H]
   \centering
     \includegraphics[clip,width=10.0cm]{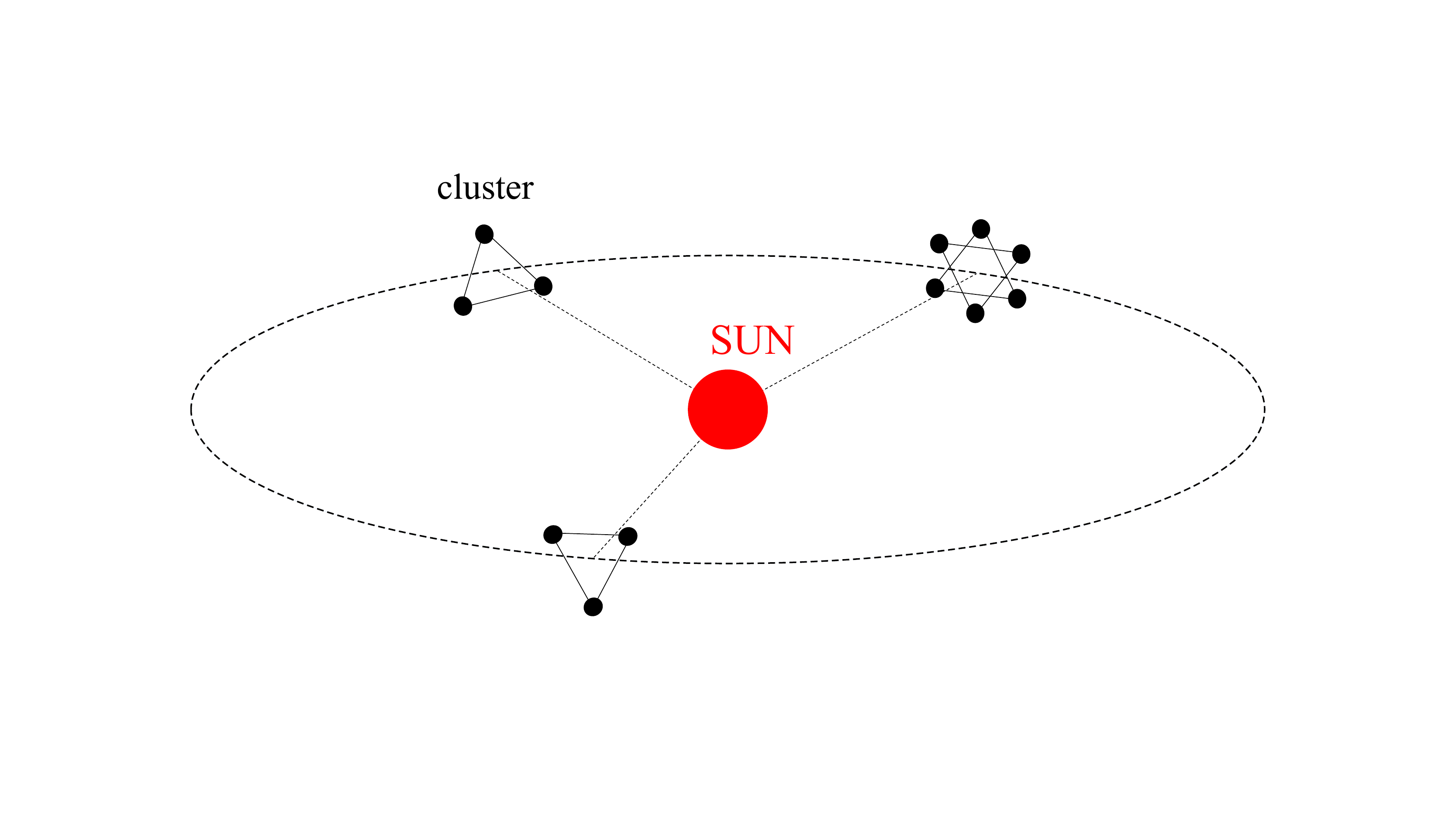}
   \caption{Schematic geometry of DECIGO. It consists of four clusters along the heliocentric orbit of the earth. Each cluster is composed of three differential FP Michelson interferometers arranged in an equilateral triangle. Two clusters are placed in the same position for detecting the primordial GWs, and the other two spaced along the orbit to enhance the angular resolution.}
   \label{fig:1-1}
 \end{figure}

To detect the primordial GWs, we designed the DECi-hertz Interferometer Gravitational-wave Observatory (DECIGO) \cite{DECIGO}. It is the future Japanese gravitational wave detector in outer space, with the geometry shown in Figure~\ref{fig:1-1}. DECIGO consists of four clusters operated along the heliocentric orbit of the earth, which is the same orbit as LISA \cite{LISA}. Two of the clusters are placed in the same position to identify the primordial GWs, while the other two are placed separately to enhance the angular resolution for discrete sources.

Each cluster is composed of three drag-free satellites. Inside each satellite, two mirrors are floating. Using them as cavity mirrors, we obtain three differential Fabry-Perot (FP) Michelson interferometers with \SI{60}{degrees} between its two arms. Sharing arms in each interferometer is possible due to the cluster's shape, an equilateral triangle. DECIGO detects GWs by measuring a change in the cavity length caused by GWs. Having very long FP cavities in outer space, DECIGO can detect GWs mainly between \SI{0.1}{Hz}$-$\SI{10}{Hz}. As for the laser light, each laser source in each satellite is independent.

We set several parameters used in the original DECIGO proposal \cite{DECIGO2}: mirror radius $R=$ \SI{0.5}{m}, cavity length $L=$ \SI{1000}{km}, finesse $\mathcal{F}=$ \SI{10}{}, laser wavelength $\lambda=$ \SI{515}{nm}, and laser power $P_0=$ \SI{10}{W}. These parameters, together with the correlation of the two clusters in the same position, were employed to provide a good target sensitivity to detect the primordial GWs assuming $\Omega_\mathrm{gw} = 2\times10^{-15}$. However, since the original design study, the upper limit of the primordial GWs was updated to be $\Omega_\mathrm{gw} = 1\times10^{-16}$ from observations by the Planck satellite \cite{Planck}, and other electro-magnetic observations \cite{Kuroyanagi}.

This motivated us to improve the target sensitivity \cite{Yamada}. In this paper, we optimize various parameters for the quantum noise of the detector as a function of cavity mirror radius $R$: cavity length $L$, mirror reflectivity $r$, and laser power $P_0$, including the effects of diffraction loss of the light. While the diffraction loss is negligible in the sensitivity design of the ground-based detectors, it is critical in the DECIGO design because DECIGO uses long cavity lengths constructed from finite-size mirrors. Although in this work we consider only quantum noise, in the actual design of DECIGO we must also consider other noise sources such as thermal noise and gravity gradient noise. Still, once we establish the method to optimize the parameters for the quantum noise, we can easily do the optimization by including other noise sources in this method.

\section{Theory}
\begin{figure}[htbp]
   \centering
     \includegraphics[clip,width=10.0cm]{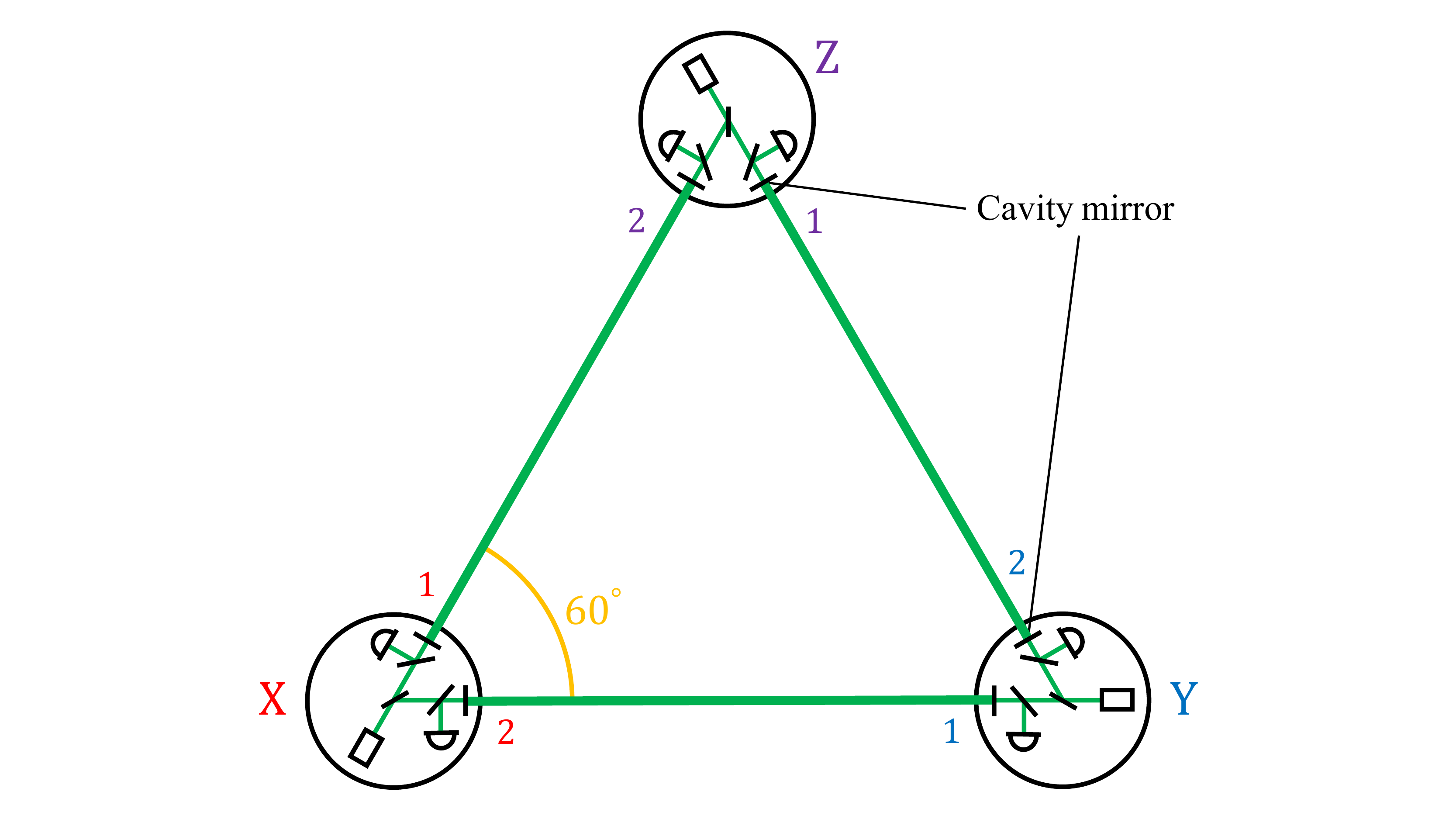}
   \caption{Schematic configuration of one cluster in DECIGO. $X,Y,Z$ denote the three interferometers in a cluster. Also, $1,2$ represent the arms in each interferometer.}
   \label{fig:2-1}
 \end{figure}

Figure~\ref{fig:2-1} shows the configuration of one cluster in DECIGO shown in Figure~\ref{fig:1-1}. As mentioned in the previous section, it has two unique characteristics:
   \begin{inparaenum}[(1)]
     \item each interferometer has \SI{60}{degrees} between its arms,
     \item each interferometer shares one arm with the other two interferometers.
   \end{inparaenum}
 Thus, the GW signal and the noise from one cluster in DECIGO should be considered properly.

In order to optimize the signal-to-noise ratio (SNR) in DECIGO, first, we have to calculate the power spectrum density (PSD) of the quantum noise of one cluster, $S_\mathrm{h}^\mathrm{cluster} (f)$, and compare it with that of the primordial GWs strain $S_\mathrm{h} (f)$. The strain spectrum is related to the normalized energy density $\Omega_\mathrm{gw} (f)$ in~\cite{Mingarelli}:
  \begin{equation}
      \Omega_\mathrm{gw} (f) = \frac{2 \pi^2}{3 {H_0}^2} f^3 S_\mathrm{h} (f)~,
    \label{eq:2-1}
  \end{equation}
where $H_0$ represents Hubble constant, and its value is \SI{70}{km/s/Mpc}.

First, we calculate one triangular detector sensitivity, $S_\mathrm{h}^\mathrm{cluster} (f)$, starting with the quantum noise of one differential FP interferometer. Then, we calculate SNR for two correlated clusters. This necessity is based on the fact that one cluster cannot detect the primordial GWs having stationary, isotropic, and non-polarized waves. In other words, for detection, we have to employ at least two clusters with correlation. Hence, we obtain the total SNR with $S_\mathrm{h} (f)$.

After that, we optimize the DECIGO parameters by determining the maximum SNR for a given cavity mirror radius $R$, and it enables to optimize the target sensitivity. In this paper, our purpose is to optimize the sensitivity of DECIGO for the detection of the primordial GWs: we concentrate on the two clusters in the same position.

There are two subsections. First, we concentrate on the quantum noise PSD of one cluster in DECIGO. Next, we derive the formula of the total SNR and optimize parameters, such as cavity length $L$, mirror reflectivity $r$, and laser power $P_0$.

\subsection{The Formula of PSD for one cluster in DECIGO $S_\mathrm{h}^\mathrm{cluster} (f)$}
At interferometer $i~(= X,Y,Z)$ in Figure~\ref{fig:2-1}, we can obtain the interferometer output from the GW strain and the noise:
  \begin{equation}
      s_i (t) = h_i (t) + n_i(t)~,
    \label{eq:2-a-1}
  \end{equation}
where $h_i(t)$ is the interferometer output caused by the GW strain and $n_i(t)$ represents the noise in the interferometer. The noise in each interferometer is correlated with the other two interferometers, $(X,Y)$, $(Y,Z)$, and $(Z,X)$, because each interferometer shares one arm with each of the other interferometers. When we consider the noise matrix of three interferometers $X,Y,Z$, it has two characteristics based on the unique shape of a cluster, the equilateral triangle. One is that the correlation matrix is symmetric because of the completely symmetric shape of one cluster. The other one is that its diagonal components are the same and its off-diagonal components are also the same since each interferometer has identical configuration. Thus, we define its diagonal components as $P_\mathrm{d}$ and off-diagonal components as $P_\mathrm{o}$, and the noise matrix of three interferometers in one cluster is written in Eq.(\ref{eq:2-a-2}) \cite{Prince}:
  \begin{equation}
      \left( \begin{array}{ccc}
        P_\mathrm{d} & P_\mathrm{o} & P_\mathrm{o} \\
        P_\mathrm{o} & P_\mathrm{d} & P_\mathrm{o} \\
        P_\mathrm{o} & P_\mathrm{o} & P_\mathrm{d}
      \end{array} \right)
    \label{eq:2-a-2}
  \end{equation}
Note that $P_\mathrm{d}$ and $P_\mathrm{o}$ are expressed in terms of the noise signal $n_i$:
  \begin{numcases}{}
      P_\mathrm{d} \equiv \langle n_i(f) {n_i(f)}^* \rangle~, \label{eq:2-a-3a} \\
      P_\mathrm{o} \equiv \langle n_i(f) {n_j(f)}^* \rangle~, \label{eq:2-a-3b}
  \end{numcases}
where the indices $i,j$ take the values $X,Y,Z$, and $i$ differs from $j$.

\begin{figure}[htbp]
    \centering
      \includegraphics[clip,width=10.0cm]{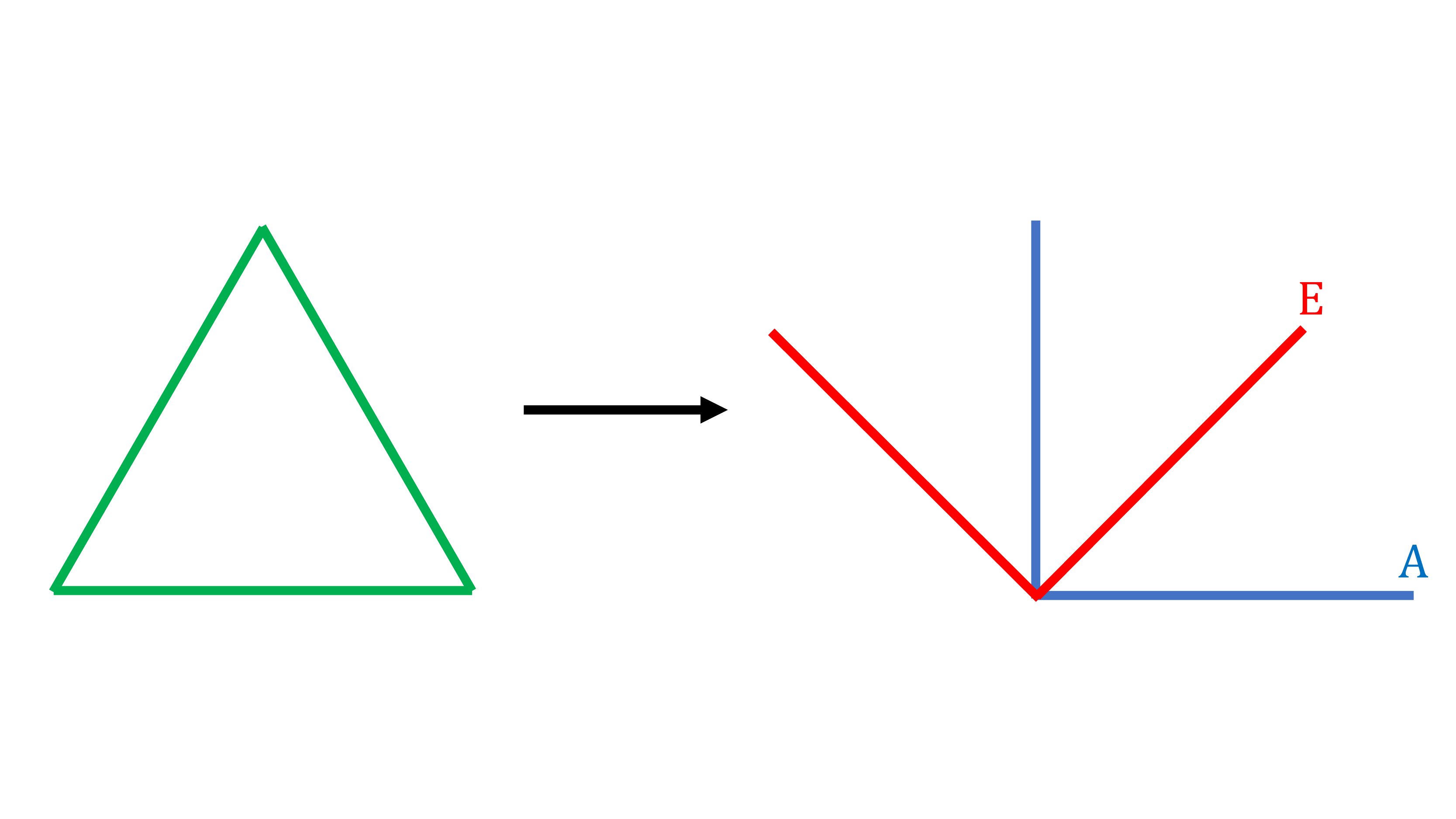}
    \caption{Transition from one cluster to two effective interferometers. The blue one shows the $A$-mode, and the red one shows the $E$-mode. The effective interferometers have \SI{90}{degrees} arm angle, and the $E$-mode is rotated by \SI{45}{degrees} from the $A$-mode.}
    \label{fig:2-a-1}
  \end{figure}

In Eq.(\ref{eq:2-a-2}), the correlation between interferometers causes difficulty in calculating the appropriate noise PSD of a cluster. Thus, we diagonalize the noise matrix. As a result, we obtain the diagonalized linear combination of three interferometers:
  \begin{equation}
      A = \frac{X - Y}{\sqrt{2}}~,~~E = \frac{X + Y - 2Z}{\sqrt{6}}~,~~T = \frac{X + Y + Z}{\sqrt{3}}~.
    \label{eq:2-a-4}
  \end{equation}
Their eigenvalues are $P_\mathrm{A} = P_\mathrm{d} - P_\mathrm{o}$, $P_\mathrm{E} = P_\mathrm{d} - P_\mathrm{o}$, $P_\mathrm{T} = P_\mathrm{d} + 2P_\mathrm{o}$. Among these combinations, however, the $T$-mode cannot be utilized because the GW signal vanishes in the summing strain data from each interferometer, at low frequencies $f < f_\mathrm{p}$ where $f_\mathrm{p}$ is the cavity pole frequency. Hence, we concentrate on two modes: the $A$-mode and the $E$-mode in Eq.(\ref{eq:2-a-4}). Figure~\ref{fig:2-a-1} represents the shape of two modes, and they are effectively right angle interferometers, with the $E$-mode interferometer rotated by \SI{45}{degrees} from the $A$-mode interferometer.

We evaluate the relationship of the PSD of GWs between interferometer $X$ and the linear combination of three interferometers $A$. We define two angular parameters, $(\theta, \phi)$: $\theta$ as the zenith angle with $z$ axis which is perpendicular to the cluster plane, and $\phi$ as the azimuth angle on the cluster plane.
We also define angular parameter $(\theta_i, \phi_i)$ in each interferometer $i$ by rotating $\phi$ around $z$ axis, that is, $\theta_i$ is equal to $\theta$, and each $\phi_i$ is different by \SI{120}{degrees} in the DECIGO's case, as shown in Figure~\ref{fig:2-1}. In addition, we assume the GW polarization angle $\psi$. Under this definition, we estimate the strain signal of GWs in interferometers $X$ and $Y$:
  \begin{numcases}{}
      h_\mathrm{X} \left( f; \theta_\mathrm{X}, \phi_\mathrm{X}, \psi; \beta \right) = \sum_{a = +,\times} F^{a} \left( \theta, \phi, \psi; \beta \right) h_{a} (f)~, \label{eq:2-a-5a} \\
      h_\mathrm{Y} \left( f; \theta_\mathrm{Y}, \phi_\mathrm{Y}, \psi; \beta \right) = \sum_{a = +,\times} F^{a} \left( \theta, \phi + 2\pi/3, \psi; \beta \right) h_{a} (f)~. \label{eq:2-a-5b}
  \end{numcases}
Note that $\beta$ is the angle between two arms in one interferometer and is equivalent to \SI{60}{degrees} in DECIGO. Besides, $F^{a},~(a = +,\times)$ are directional dependence of plus-mode and cross-mode GWs, respectively~\cite{PGWs}. We calculate the PSD of the GWs for interferometer $X$, $S_\mathrm{X}(f)$, and for the combination $A$, $S_\mathrm{A}(f)$ employing Eq.(\ref{eq:2-a-4}-\ref{eq:2-a-5b}). Then we compare $S_\mathrm{A}(f)$ with $S_\mathrm{X}(f)$:
  \begin{equation}
      S_\mathrm{A}(f) = \frac{3}{2} S_\mathrm{X}(f)~.
    \label{eq:2-a-6}
  \end{equation}
The noise PSD for the $A$-mode is the same as its eigenvalue $P_\mathrm{A}$. Therefore, the SNR of the $A$-mode with $S_\mathrm{X}(f)$ is written in the form
  \begin{equation}
      \mathrm{SNR} = \left[ \int_{f_\mathrm{min}}^{f_\mathrm{max}} \frac{S_\mathrm{A} (f)}{P_\mathrm{A} (f)} \mathrm{d} f \right]^{1/2} = \left[ \int_{f_\mathrm{min}}^{f_\mathrm{max}} \frac{S_\mathrm{X} (f)}{\frac{2}{3} \left(  P_\mathrm{d} - P_\mathrm{o} \right)} \mathrm{d} f \right]^{1/2}~.
    \label{eq:2-a-7}
  \end{equation}
Note that $f_\mathrm{max}$ is less than $f_\mathrm{p}$. Consequently, the noise PSD for the $A$-mode, $S_\mathrm{gw}^\mathrm{A}(f)$, with $S_\mathrm{X}(f)$ is:
  \begin{equation}
      S_\mathrm{gw}^\mathrm{A}(f) = \frac{2}{3} \left(  P_\mathrm{d} - P_\mathrm{o} \right)~.
    \label{eq:2-a-8}
  \end{equation}
We use the same method to evaluate the relationship between interferometers $Y,Z$ and the linear combination $A$, and between $X,Y,Z$ and $E$. This gives the same value as the noise PSD for the $A$-mode with $S_i(f)$ and for the $E$-mode with $S_i(f)$ in Eq.(\ref{eq:2-a-8}).

Regarding the PSD of GWs for each interferometer in a cluster, each interferometer $X, Y$ and $Z$ detects the primordial GWs with a PSD of GW signal in the interferometer $S_\mathrm{gw} (f)$ of
  \begin{equation}
      S_\mathrm{gw} (f) = \frac{\sin^2 \beta}{5} S_\mathrm{h} (f)~,
    \label{eq:2-3}
  \end{equation}
in the whole sky average~\cite{Mingarelli}. That is, the PSD of GWs for interferometer $i~(= X,Y,Z)$, $S_i(f)$, is equivalent to $S_\mathrm{gw} (f)$. Thus, the noise PSD for the combination $A$, ${S}_\mathrm{h}^\mathrm{A}(f)$, with $S_\mathrm{h}(f)$ is obtained by imposing Eq.(\ref{eq:2-3}):
  \begin{equation}
      S_\mathrm{h}^\mathrm{A}(f) = \frac{10}{3 \sin^2 \beta} \left( P_\mathrm{d} - P_\mathrm{o} \right)~.
    \label{eq:2-a-9}
  \end{equation}

\begin{figure}[H]
  \centering
    \includegraphics[clip,width=8.0cm]{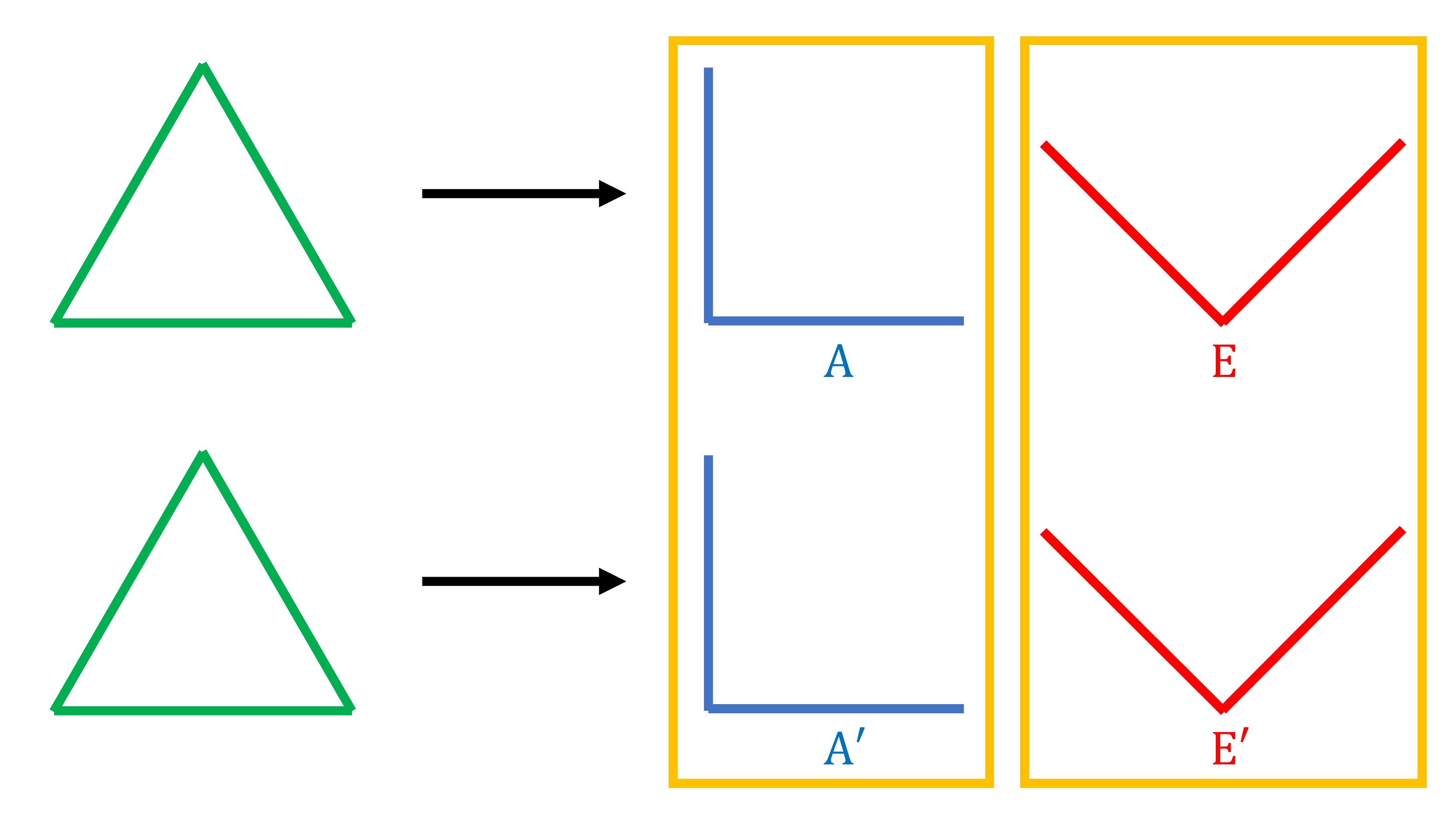}
  \caption{Pairs of the $L$ shape interferometers. The $E$-mode is rotated by \SI{45}{degrees} from the $A$-mode, shown in Figure~\ref{fig:2-a-1}. Besides, every GW mode is generated by a linear combination: the plus-mode and the cross-mode. In other words, the pairs $(A,E')$ and $(A',E)$ are not correlated.}
  \label{fig:2-a-2}
\end{figure}

Two linear combinations, $A$ and $E$, are derived individually from each cluster, which are then correlated in order to detect the primordial GWs. We label these combinations obtained from each cluster as $(A,E)$, and $(A',E')$, respectively. As shown in Figure~\ref{fig:2-a-2}, only the $AA'$-pair and $EE'$-pair have correlations since two effective interferometers are rotated by \SI{45}{degrees} to each other in Figure~\ref{fig:2-a-1}. Thus, we have to consider the number of pairs in Eq.(\ref{eq:2-a-9}), and we obtain the PSD for one cluster in DECIGO $S_\mathrm{h}^\mathrm{cluster}(f)$ with the PSD $S_\mathrm{h} (f)$:
  \begin{equation}
      S_\mathrm{h}^\mathrm{cluster}(f) = \frac{5\sqrt{2}}{3\sin^2 \beta}\left(  P_\mathrm{d} - P_\mathrm{o} \right)~.
    \label{eq:2-a-10}
  \end{equation}
Note that the factor of $\sqrt{2}$ improvement in the sensitivity comes from the fact that the added noise in a factor $\sqrt{2}$ larger while the signal increases by a factor of 2.

Eq.(\ref{eq:2-a-10}) is not expressed in terms of the quantum noise: shot noise $h_\mathrm{shot} (f)$ and radiation pressure noise $h_\mathrm{rad}(f)$. Therefore, we rewrite $S_\mathrm{h}^\mathrm{cluster}(f)$ in terms of two kinds of quantum noise. For the sake of simplicity, we concentrate on the quantum noise; we eliminate other noise sources. The formulae of the quantum noise $h_\mathrm{shot} (f)$ and $h_\mathrm{rad} (f)$ for each FP interferometer in one cluster with diffraction loss in~\cite{Iwaguchi} are:
 \begin{numcases}{}
     h_\mathrm{shot} (f) = \frac{1}{4 \pi L} \frac{ \left(1 - {r_\mathrm{eff}}^2 \right)^2}{{t_\mathrm{eff}} (tD) r_\mathrm{eff}} \sqrt{\frac{4 \pi \hbar c \lambda}{P_0}} \sqrt{1 + \left( \frac{f}{f_\mathrm{p}} \right)^2}~, \label{eq:shot} \\
     h_\mathrm{rad} (f) = \frac{4}{m L \left( 2 \pi f \right)^2} \frac{{t_\mathrm{eff}}^2 (rD)^2 (1 + {r_\mathrm{eff}}^2)}{(1 - {r_\mathrm{eff}}^2)^2} \sqrt{\frac{\pi \hbar P_0}{c \lambda}} \sqrt{\frac{1}{1 + \left( f / f_\mathrm{p} \right)^2}}~, \label{eq:rad}
 \end{numcases}
and the parameters are defined in Table~\ref{tab:2-1}. Note that $D$ is an effect of diffraction loss, which is defined later. The case of the general differential FP interferometer is discussed in~\cite{Iwaguchi}, thus we derive Eq.(\ref{eq:shot}) and Eq.(\ref{eq:rad}) with the assumption of the DECIGO settings: input and end mirror have identical mirror radius, curvature radius, and reflectivity. Also, we only consider low frequencies $f < f_\mathrm{p}$ because two kinds of noise are approximated at high frequencies $f > f_\mathrm{p}$ in~\cite{Iwaguchi}.

\begin{table}[H]
  \centering
  \caption{Definition of the DECIGO parameters.}
  \label{tab:2-1}
    \begin{tabular}{cc}
      \Hline
        \textbf{Symbol} & \textbf{Parameter} \\ \hline
        $m$ & Cavity mirror mass \\
        $L$ & Cavity length \\
        $P_0$ & Laser power \\
        $\lambda$ & Laser wavelength \\ \hline
        $r_\mathrm{eff} \equiv r D^2$ & Effective mirror reflectivity \\
        $t_\mathrm{eff} \equiv t D^2$ & Effective mirror transmissivity \\
        $r$ & Real mirror reflectivity \\
        $t$ & Real mirror transmissivity \\
        $D$ & Effect of diffraction loss \\ \hline
        $f_\mathrm{p} \equiv c / 4 \mathcal{F}_\mathrm{eff} L$ & Cavity pole frequency \\
        $\mathcal{F}_\mathrm{eff} \equiv \pi r_\mathrm{eff}/\left( 1 - {r_\mathrm{eff}}^2 \right)$ & Effective finesse \\
      \Hline
    \end{tabular}
\end{table}

On the other hand, we also define the noise strain data in each interferometer in Figure~\ref{fig:2-1}. Shot noise is the sensor noise caused by the fluctuations of photon numbers at photodetector (PD) in each differential FP interferometer, and is set as
  \begin{equation}
      n_{\mathrm{shot},i\alpha}(f)~,~~(i = X,Y,Z,~\alpha = 1,2)~,
    \label{eq:2-a-11}
  \end{equation}
where $\alpha$ is the index of arm in each interferometer, as shown in Figure~\ref{fig:2-1}. Besides, radiation pressure noise is the displacement noise that occurred at FP cavity mirrors in each arm. It also is caused by each laser source in each interferometer. Therefore, we set it for every FP cavity arm derived from each interferometer as follows:
  \begin{equation}
      n_{\mathrm{rad},i\alpha} (f)~.
    \label{eq:2-a-12}
  \end{equation}
Every $n_{\mathrm{shot},i\alpha}(f)$ is independent, and every $n_{\mathrm{rad},i\alpha}(f)$ is also independent. Besides, $n_{\mathrm{shot},i\alpha}$ only have its correlations with that in the same Interferometer. Consequently, the relation between $n_{\mathrm{shot},i\alpha}(f)$ and $h_\mathrm{shot} (f)$ and between $n_{\mathrm{rad},i\alpha}(f)$ and $h_\mathrm{rad} (f)$ are given by
  \begin{numcases}{}
      \sqrt{{n_{\mathrm{shot},i1}(f)}^2 + {n_{\mathrm{shot},i2}(f)}^2} = h_\mathrm{shot} (f)~, \label{eq:2-a-13a} \\
      \sqrt{{n_{\mathrm{rad},i1}(f)}^2 + {n_{\mathrm{rad},i2}(f)}^2} = \sqrt{{n_{\mathrm{rad},i1}(f)}^2 + {n_{\mathrm{rad},j2}(f)}^2} = h_\mathrm{rad} (f)~. \label{eq:2-a-13b}
  \end{numcases}
Employing these relations, we rewrite Eq.(\ref{eq:2-a-10}) with $h_\mathrm{shot} (f)$ and $h_\mathrm{rad} (f)$.

First, we concentrate on $P_\mathrm{d}$, specifically, that of interferometer $X$; $P_\mathrm{d} = \langle n_\mathrm{X}(f) {n_\mathrm{X}(f)}^* \rangle$. In Figure~\ref{fig:2-1}, it includes shot noise that occurred from interferometer $X$ only and four different sources of  radiation pressure noise. Each radiation pressure noise is derived from \begin{inparaenum}[(1)]
  \item interferometer $X$, arm 1,
  \item interferometer $X$, arm 2,
  \item interferometer $Y$, arm 1,
  \item interferometer $Z$, arm 2.
\end{inparaenum}
The latter two are contained because of arm sharing. Consequently, $P_\mathrm{d}$ for interferometer $X$ is
  \begin{align}
      P_\mathrm{d} &= \langle n_\mathrm{X}(f) {n_\mathrm{X}(f)}^* \rangle \notag \\
          &= {n_\mathrm{shot,X1}(f)}^2 + {n_\mathrm{shot,X2}(f)}^2 + {n_\mathrm{rad,X1}(f)}^2 + {n_\mathrm{rad,X2}(f)}^2 + {n_\mathrm{rad,Y1}(f)}^2 + {n_\mathrm{rad,Z2}(f)}^2 \notag \\
          &= {h_\mathrm{shot}(f)}^2 + 2 {h_\mathrm{rad}(f)}^2~.
    \label{eq:2-a-14}
  \end{align}

Next, we concentrate on $P_\mathrm{o}$, specifically, between interferometer $X$ and $Y$; $P_\mathrm{o} = \langle n_\mathrm{X}(f) {n_\mathrm{Y}(f)}^* \rangle$. Shot noise has no correlation between the PD in different interferometer, that is, its value is $0$. On the other hand, regarding radiation pressure noise, two sources associated in arm sharing exist, \begin{inparaenum}[(1)]
  \item interferometer $X$, arm 2,
  \item interferometer $Y$, arm 1,
\end{inparaenum}
as shown in Figure~\ref{fig:2-1}. Hence, $P_\mathrm{o}$ in this case is
  \begin{align}
      P_\mathrm{o} &= \langle n_\mathrm{X}(f) {n_\mathrm{Y}(f)}^* \rangle \notag \\
          &= {n_\mathrm{rad,X2}(f)}^2 + {n_\mathrm{rad,Y1}(f)}^2 \notag \\
          &= {h_\mathrm{rad}(f)}^2~.
    \label{eq:2-a-15}
  \end{align}
We evaluate $P_\mathrm{d}$, $P_\mathrm{o}$ with other noise combinations employing the previous method and obtain the same result.

Finally, we substitute Eq.(\ref{eq:2-a-14}) for $P_\mathrm{d}$ and Eq.(\ref{eq:2-a-15}) for $P_\mathrm{o}$, and rewrite Eq.(\ref{eq:2-a-10}) :
  \begin{equation}
      S_\mathrm{h}^\mathrm{cluster}(f) = \frac{5\sqrt{2}}{3\sin^2\beta} \left( {h_\mathrm{shot}(f)}^2 + {h_\mathrm{rad}(f)}^2 \right)~.
    \label{eq:2-a-16}
  \end{equation}
This equation represents the following characteristics. Each interferometer with $\beta$ arm angle has the particular GW signal in Eq.(\ref{eq:2-3}). It introduces the factor $5/\sin^2\beta$. Also, three interferometers contained in one cluster cause another factor of one third. Finally, arm sharing impairs the factor by a further factor of $\sqrt{2}$.

\subsection{Optimization of the DECIGO parameters}
To optimize the DECIGO parameters, we calculate the SNR of two clusters in DECIGO. As is mentioned above, we cannot detect the primordial GWs with one cluster since they are steady, isotropic, and non-polarized waves. Instead, we have to utilize the correlations between two clusters to detect the primordial GWs.

At low frequencies $f < f_\mathrm{p}$, GWs remain in the same phase while the light is bounced back and forth in the FP cavity. The SNR in DECIGO increases with increased observation time; the SNR of the correlated signal from two clusters is enhanced by its observation time.

The SNR with the correlation between each cluster in~\cite{Courty} is written:
  \begin{equation}
      \mathrm{SNR} = \frac{3 {H_0}^2}{10 \pi^2} \sqrt{T_\mathrm{obs}} \left[ \int_{0.1}^1 \frac{2 {\gamma^2(f)} {\Omega^2_\mathrm{gw} (f)}}{f^6 P_1 (f) P_2 (f)} \mathrm{d} f \right]^{1/2}~.
    \label{eq:2-11}
  \end{equation}
Note that $P_j (f),~(j = 1,2)$ is the PSD of each cluster, and we assume $P_1 (f) = P_2 (f) = S_\mathrm{h}^\mathrm{cluster} (f)/5$ to eliminate whole sky average redundancy. Formally, the formula includes $\gamma (f)$, the normalized overlap reduction function, equivalent to 1, because the two cluster's antenna patterns from the primordial GWs are identical, despite their opposite orientation in the same plane in Figure~\ref{fig:1-1}. In the estimates of sensitivity below, we assume that observation time $T_\mathrm{obs}$ is three years. Regarding the frequency range, the confusion limiting noise from white dwarf binaries prevents DECIGO from detecting the primordial GWs below \SI{0.1}{Hz}. Thus, we calculate the SNR from \SI{0.1}{Hz} to \SI{1}{Hz} to optimize the sensitivity around \SI{0.1}{Hz}.

\begin{figure}[htbp]
   \centering
     \includegraphics[clip,width=10.0cm]{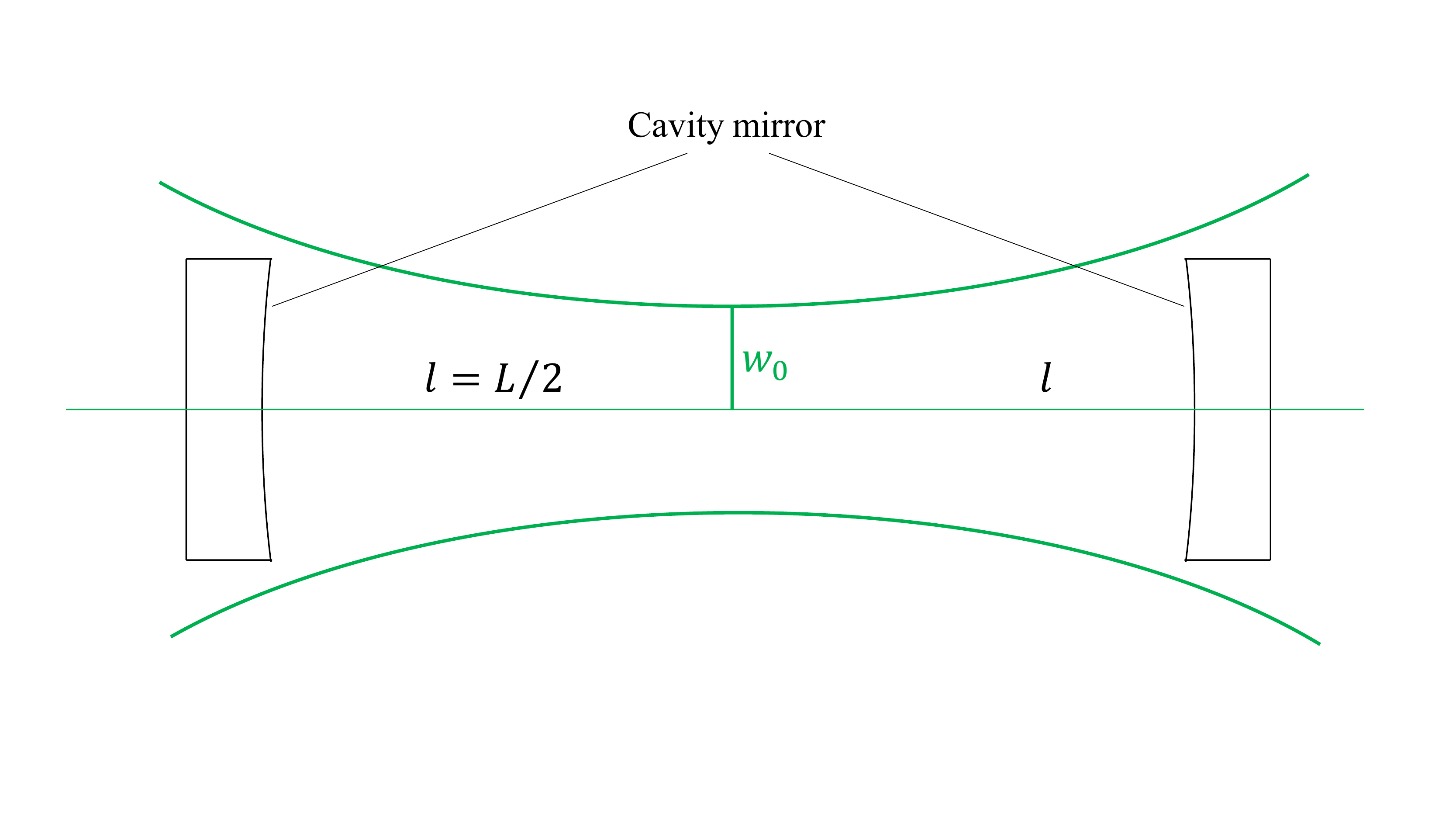}
   \caption{Detailed configuration of a FP cavity in DECIGO. Each FP cavity is shared by two interferometers. The input and end mirror are identical, that is, this FP cavity is symmetrical with respect to its beam waist point. The green horizontal line represents the optical axis of laser light. Also, the green curve shows the light spreading.}
   \label{fig:2-3}
 \end{figure}

Figure~\ref{fig:2-3} shows the configuration of a typical FP cavity in each interferometer's arm. We place two mirrors separated by the cavity length $L$, and each mirror possesses its radius $R$ and its radius of curvature. Each mirror is located at a distance of $l = L/2$ from the beam waist position at each end of the cavity. We also define $z_\mathrm{R}$ as the Rayleigh length of the laser beam.

Inside the FP cavity, the beam size of the laser light entering from the input mirror decreases toward the beam waist, and increases on the way from the beam waist to the end mirror. At the input and end mirror, a part of the light power is lost if the mirror radius is smaller than that of the beam size; a small diffraction loss occurs. Thus, the mirror effective reflectivity with this loss $r_\mathrm{eff}$ is smaller than the actual reflectivity of the mirror itself $r$:
   \begin{equation}
       r_\mathrm{eff} \equiv r \times D^2~,
     \label{eq:2-12}
   \end{equation}
 where $D$ is the effect of diffraction mentioned in the previous subsection. In Eq.(\ref{eq:2-12}), $r$ is multiplied by the squared $D$ because we consider two effects: leakage loss and higher-order mode loss~\cite{Iwaguchi}. The leakage loss is imposed when a part of laser power is lost due to a finite mirror radius, and the higher-order mode loss is considered because the FP cavity is adjusted to the resonance state for the fundamental mode of the laser light. It decreases with the increase of diffraction. Using parameters $R$, $l~(\text{or}~L)$, $z_\mathrm{R}$ and $\lambda$, we can rewrite Eq.(\ref{eq:2-12}) as
   \begin{equation}
       r_\mathrm{eff} = r \left( 1 - \mathrm{exp} \left[ - \frac{2 \pi z_\mathrm{R}}{\lambda \left( l^2 + {z_\mathrm{R}}^2 \right)} R^2 \right] \right)~.
     \label{eq:2-13}
   \end{equation}
 Hence, $D^2$ is represented:
   \begin{equation}
       D^2 = 1 - \mathrm{exp} \left[ - \frac{2 \pi z_\mathrm{R}}{\lambda \left( l^2 + {z_\mathrm{R}}^2 \right)} R^2 \right]~.
     \label{eq:2-14}
   \end{equation}
 Per Eq.(\ref{eq:2-14}), $D^2$ ranges from \SI{0}{} to \SI{1}{}.

\begin{figure}[H]
     \centering
       \includegraphics[clip,width=8.0cm]{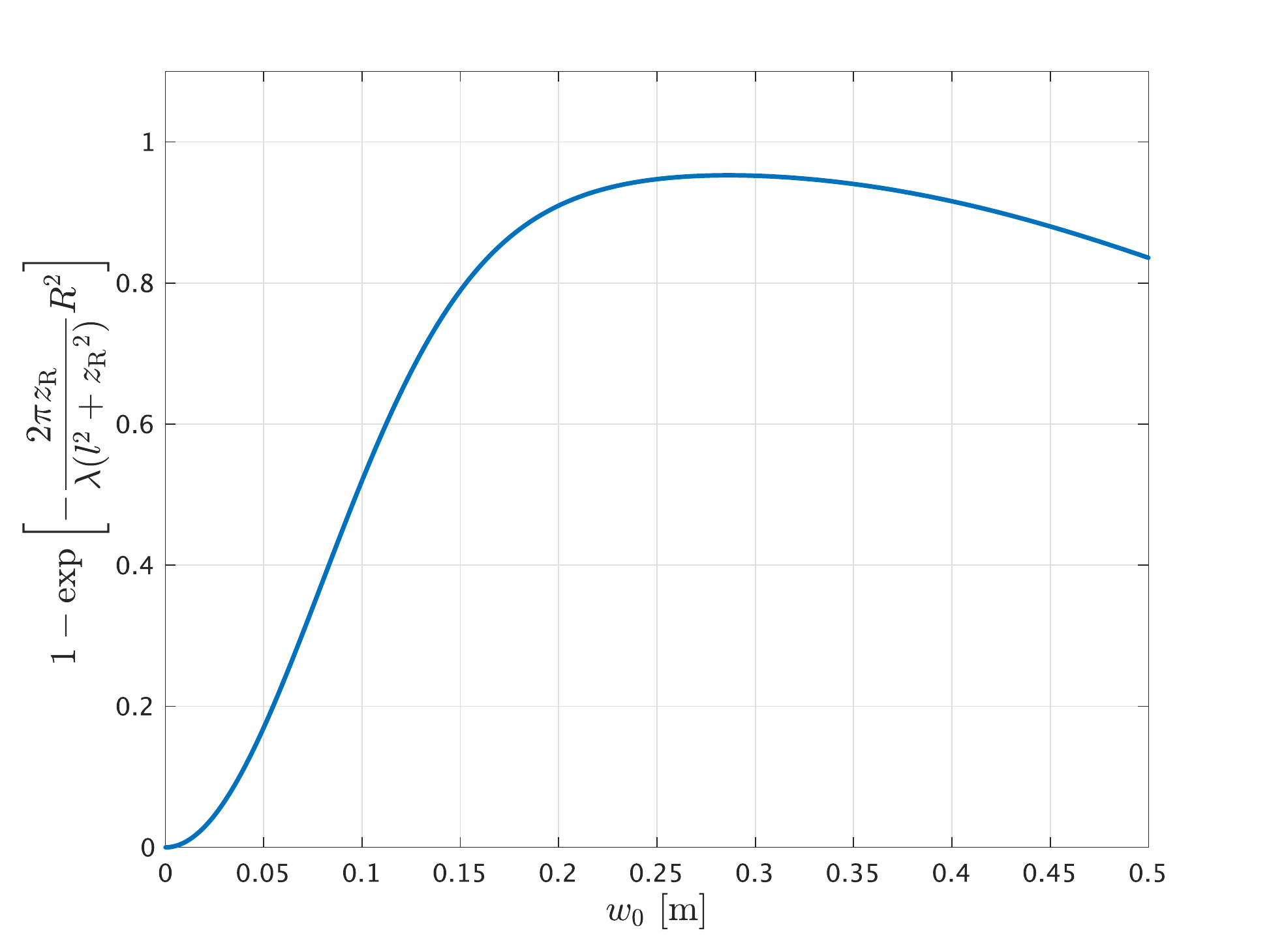}
     \caption{Dependence of $D^2$ on $w_0$ with the default DECIGO parameters. There is a maximum $D^2$ for specific beam waist $w_0$.}
     \label{fig:2-4}
   \end{figure}

To determine the appropriate parameters maximizing the SNR of the two clusters in Eq.(\ref{eq:2-11}), first, we concentrate on $D^2$ and optimize. Figure~\ref{fig:2-4} shows the $D^2$ curve for a given beam waist $w_0$ with the default DECIGO. The radius of the beam at the FP cavity mirrors depends on the beam waist size. The beam size is large at the mirrors as a result of divergence if the beam waist is significantly small, while the beam spot is naturally large if the beam waist is significantly large. Thus, an appropriate beam waist can maximize $D^2$ for given $R$ and $L$. The beam waist $w_0$ is related to the Rayleigh length $z_\mathrm{R}$:
   \begin{equation}
       z_\mathrm{R} = \frac{\pi {w_0}^2}{\lambda}~.
     \label{eq:2-15}
   \end{equation}
 Eq.(\ref{eq:2-15}) shows $z_\mathrm{R}$ increases linearly with the square of $w_0$, that is, $D^2$ can be maximized with appropriate $z_\mathrm{R}$.

 Considering the confocal geometry of the cavity in DECIGO, we can determine $z_\mathrm{R}$ that maximize $D^2$ as the following:
   \begin{equation}
       z_\mathrm{R} = l = \frac{L}{2}~.
     \label{eq:2-18}
   \end{equation}
 Thus, we obtain the maximum ${D_\mathrm{opt}}^2$ as the minimum effect of diffraction loss:
   \begin{equation}
       {D_\mathrm{opt}}^2 = 1 - \mathrm{exp} \left[ - \frac{2 \pi}{L \lambda} R^2 \right]~.
     \label{eq:2-D}
   \end{equation}

Under this optimized effective reflectivity $r_\mathrm{eff}$, we calculate the total SNR in DECIGO, applying Eq.(\ref{eq:2-11}) as a function of $R$, $L$, $r$, and $P_0$:
  \begin{equation}
      \mathrm{SNR} = \mathrm{SNR}(R; L, r, P_0)~,
    \label{eq:2-19}
  \end{equation}
and calculate the largest SNR and $L$, $r$, and $P_0$ that give the SNR for $R$ as the only free parameter. In the case of mirror mass, we calculate the SNR for the two cases: the constant mirror-thickness case and the constant mirror-mass case, for different $R$.

\section{Result}
\begin{figure}[htbp]
     \subfigure[]{
       \includegraphics[clip,width=7.5cm]{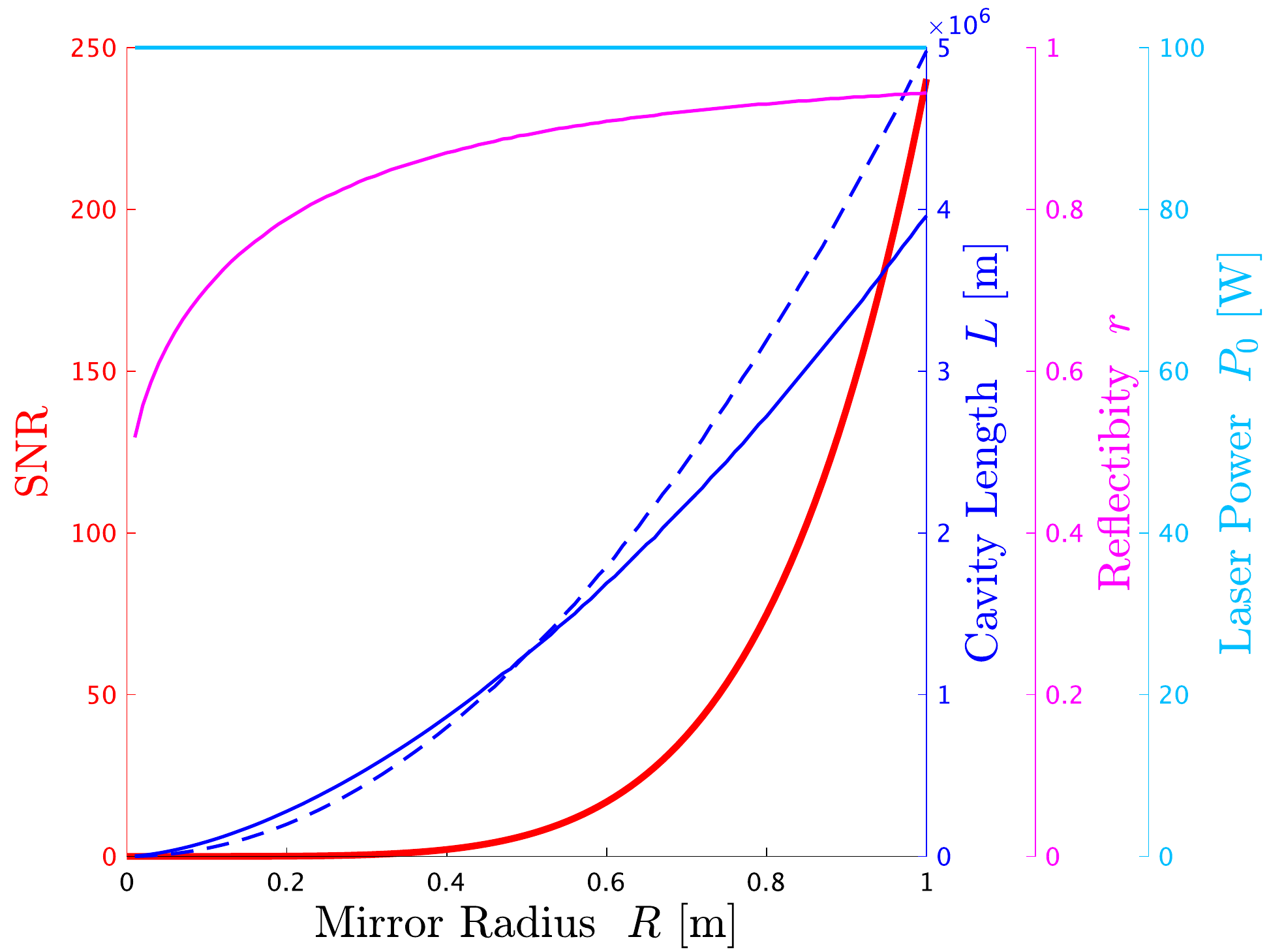}
       \label{fig:3-1-a}
     }
     \hfill
     \subfigure[]{
       \includegraphics[clip,width=7.5cm]{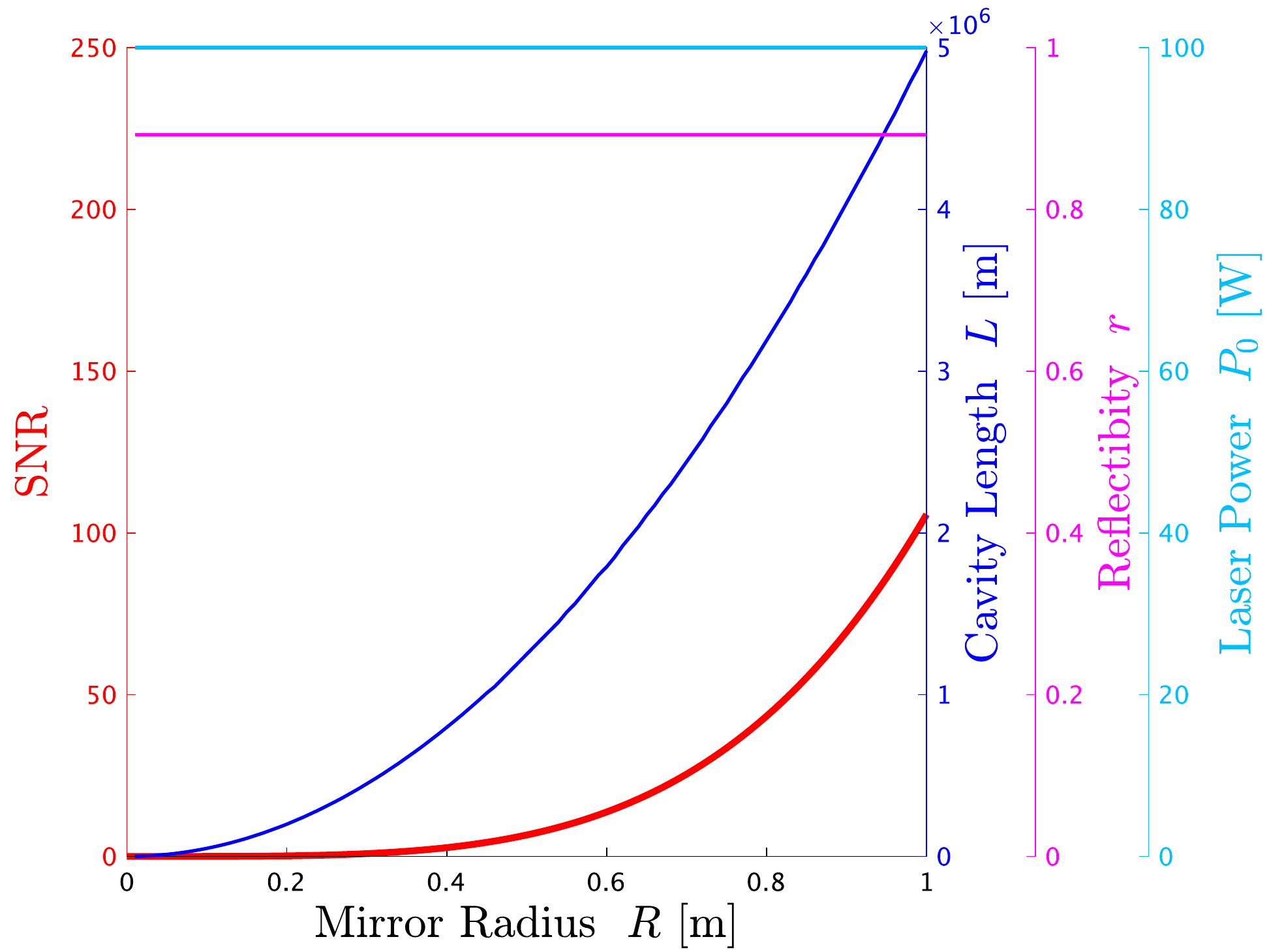}
       \label{fig:3-1-b}
     }
     \vspace{-0.5cm}
     \caption{Optimized SNR for given $R$ (red line), and the cavity length $L$ (blue line), mirror reflectivity $r$ (magenta line) and laser power $P_0$ (cyan line) which give this SNR. (a) shows the case that the mirror mass increases linearly with the squared mirror radius $R^2$ in Eq.(\ref{eq:3-1}), i.e., the constant mirror-thickness case,  and (b) shows the case of constant mirror mass: \SI{100}{kg}. Note that the dashed blue curve is the same as the solid blue curve of (b) added to compare the optimized length in both cases.}
   \label{fig:3-1}
 \end{figure}

Figure \ref{fig:3-1-a} shows the largest SNR as a function of $R$, and the parameters $L$, $r$, and $P_0$ needed to achieve the optimized SNR in the case that the mirror mass changes with $R$. Regarding the mirror mass, it increases linearly with the progression of $R^2$:
   \begin{equation}
       m = \left( \frac{R}{\SI{0.5}{m}} \right)^2 \times \SI{100}{kg}~.
     \label{eq:3-1}
   \end{equation}
 Note this assumes that the thickness of the cavity mirror is held constant as the mirror radius is increased, and that the mirror mass at $R =$ \SI{0.5}{m} is the same value of the default DECIGO setting: \SI{100}{kg}. Figure~\ref{fig:3-1-b} shows the optimized parameters in the case where the mirror mass is constant, at the default DECIGO mass.

In Figure~\ref{fig:3-1}, we calculate over limited parameters' ranges: cavity mirror radius $R$ ranging from \SI{0}{} to \SI{1}{m}, mirror reflectivity $r$ ranging from \SI{0}{} to \SI{1}{}, and laser power $P_0$ at every \SI{10}{W} from \SI{0}{} to \SI{100}{W}. In this section, we discuss the particular radius case and the free-parameter radius case.

\subsection{Same radius as that of the default DECIGO parameter}
We concentrate on the particular case: $R = $ \SI{0.5}{m}. In this case, the optimized SNR and $L$, $r$, and $P_0$ have the same results from Figure~\ref{fig:3-1-a} and Figure~\ref{fig:3-1-b} because the mirror mass in the two cases is identical. These concrete values are shown in Table~\ref{tab:3-1}. In addition, we calculate the SNR, for which the default DECIGO parameters are utilized, and its value is 3.2. Note that, in this calculation, we ignore the diffraction loss: we assume that the effect of diffraction loss $D_\mathrm{opt}$ is equal to \SI{1}{}. Compared with the default SNR, the optimized SNR is higher at the same mirror radius.
 \begin{table}[htbp]
   \centering
   \caption{Optimized SNR and parameters from Figures~\ref{fig:3-1}, $R = 0.5~\mathrm{m}$. The last line shows the effective finesse calculated from $r$ and ${D_\mathrm{opt}}^2$ in Eq.(\ref{eq:2-D}). For comparison, the parameters in default DECIGO are also listed. In default DECIGO, we show two cases: ignoring the diffraction loss and considering it. The values of $r$ and $D_\mathrm{opt}$ in the two default DECIGO cases are decided by two conditions: $\mathcal{F}_\mathrm{eff} = 10$ and $D_\mathrm{opt}$ condition. From Figure~\ref{fig:3-2-a}, the cavity pole frequencies $f_\mathrm{p}$ are a few Hz and not relevant in the frequency range where there is a chance to detect the primordial GWs.}
     \begin{tabular}{cccc}
       \Hline
         \textbf{Symbol} & \textbf{Default($D_\mathrm{opt}=1$)} & \textbf{Default($D_\mathrm{opt}\neq1$)} & \textbf{Optimized} \\ \hline
         SNR & \SI{3.2}{} & \SI{1.6}{} & \SI{6.6}{} \\
         $L$ & \SI{1000}{km} & \SI{1000}{km} & \SI{1250}{km} \\
         $r$ & \SI{0.855}{} & \SI{0.898}{} & \SI{0.892}{} \\
         $D_\mathrm{opt}$ & \SI{1}{} & \SI{0.98}{} & \SI{0.96}{} \\
         $P_0$ & \SI{10}{W} & \SI{10}{W} & \SI{100}{W} \\ \hline
         $\mathcal{F}_\mathrm{eff}$ & \SI{10}{} & \SI{10}{} & \SI{7.6}{} \\
       \Hline
     \end{tabular}
   \label{tab:3-1}
 \end{table}

 \begin{figure}[htbp]
   \subfigure[]{
     \includegraphics[clip,width=8.0cm]{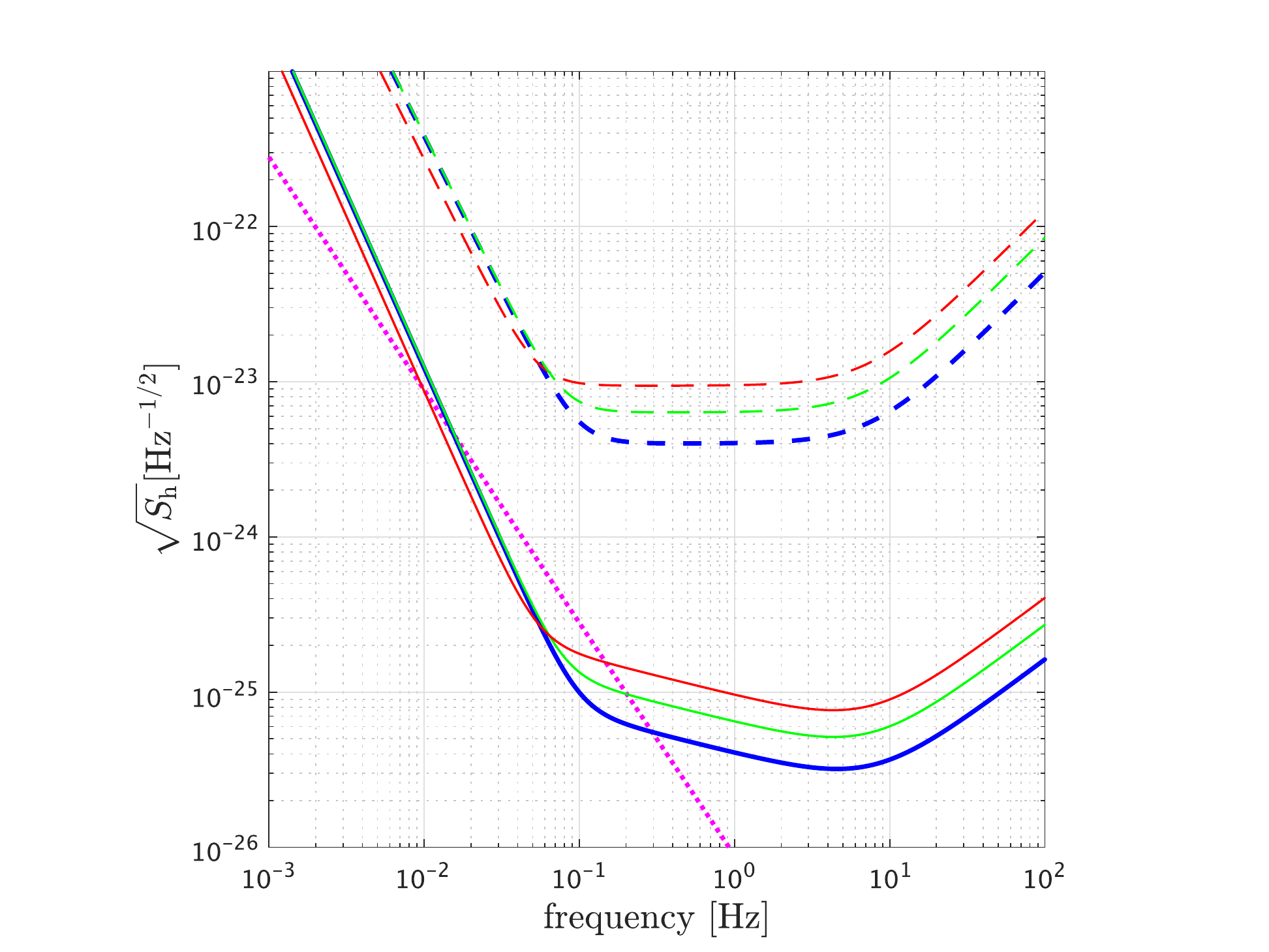}
     \label{fig:3-2-a}
   }
   \hfill
   \subfigure[]{
     \includegraphics[clip,width=8.0cm]{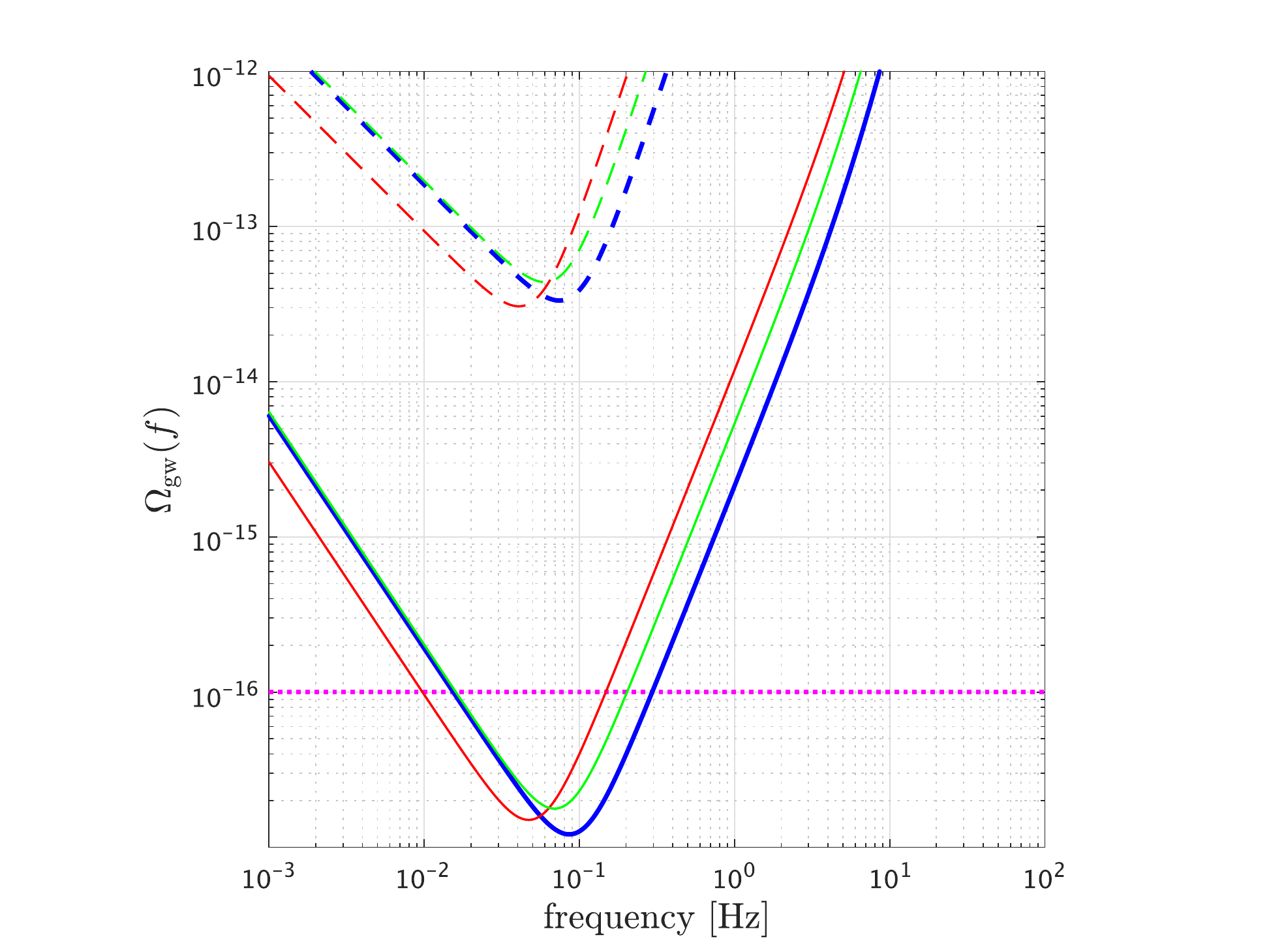}
     \label{fig:3-2-b}
   }
   \vspace{-0.5cm}
   \caption{Sensitivity curves for one cluster (dashed lines) and the correlation of two clusters in DECIGO (solid lines) in terms of strain sensitivity (a) and normalized energy density (b). The blue line shows the case for the optimized parameters in Table~\ref{tab:3-1}, the green line shows the case for the default design without the effect of the diffraction loss, and the red line shows the case for the default design with the effect of the diffraction loss. The dotted magenta line shows the primordial GWs in Eq.(\ref{eq:2-1}) with $\Omega_\mathrm{gw} = 1\times10^{-16}$.}
   \label{fig:3-2}
 \end{figure}

Figure~\ref{fig:3-2} shows three curves. The blue one is the case of optimized parameters, and the red and green ones are those of the default design, with the red curve including the effects of diffraction and the green curve ignoring it. These curves in Figure~\ref{fig:3-2-a} are drawn as the sensitivity of one cluster in DECIGO, $\sqrt{S_\mathrm{h}^\mathrm{cluster} (f)}$, and that of the correlated clusters. Also, we draw the curves as the normalized energy density, $\Omega_\mathrm{gw}(f)$, in Figure~\ref{fig:3-2-b}. Regarding the sensitivity for the correlated clusters, we have to consider two effects. The one is the observation time $T_\mathrm{obs}$ because the SNR for the correlated clusters is increased with the observation time. The other one is the times of cycles. The latter is substituted by frequency $f$. Thus, the PSD for the correlated two clusters $S_\mathrm{h}^{2\mathrm{clusters}}(f)$ is derived from Eq.(\ref{eq:2-11}) and written as
  \begin{equation}
      S_\mathrm{h}^{2\mathrm{clusters}}(f) = S_\mathrm{h}^\mathrm{cluster}(f) \frac{1}{\sqrt{T_\mathrm{obs}f}}~.
  \end{equation}
From these figures, we can see that the target sensitivity is enhanced even if we consider the diffraction loss. Considering Figure~\ref{fig:3-1}, this is because the laser power compensates for the two undesirable consequences from the other parameters.

The first one is the increase of the diffraction loss since the cavity length is longer than that of the default, $D_\mathrm{opt} \neq 1$. The longer cavity length increases the diffraction loss, i.e., $D_\mathrm{opt}$ is lowered. In Eq.(\ref{eq:2-12}), the effective reflectivity $r_\mathrm{eff}$ is also reduced. Consequently, the effective finesse $\mathcal{F}_\mathrm{eff}$ is reduced. This reduces the radiation pressure noise and increases the shot noise. The SNR is mainly limited by the shot noise at the frequency band: \SI{0.1}{Hz} - \SI{1}{Hz}; therefore, it prevents the SNR improvement.

The other one consequence is the decrease of the laser power in the FP cavity since the mirror reflectivity $r$ is smaller than that of the default, $D_\mathrm{opt} \neq 1$. This spurs on the effective finesse reduction. These two problems are solved by the laser power increasing to the limited value, and the optimized SNR is enhanced. Regarding the laser power limit, we calculate at the limited laser power range: \SI{100}{W} or less. It is based on practical considerations only.

The increase of laser power decreases the shot noise, while it causes the increase of radiation pressure noise. Thus, the laser power should be optimized to a certain value. However, this result shows the increased laser power compensates for the two effects of lowering the power insider the cavity. In other words, the SNR is improved in such a way that the cavity wastes a part of the increased laser power. Therefore, we could lengthen the cavity length and lower the mirror reflectivity if we raise the upper limit of laser power. In other words, the target sensitivity can be more enhanced.

\subsection{Mirror radius as a free parameter}
In this section, we concentrate on the case with mirror radius as a free parameter. In both cases, the optimized SNR increases with the mirror radius $R$, and the optimized $L$ also increases. This relationship is mostly because the strain sensitivity scales like $1/L$.

Other characteristics can be seen by comparing Figure~\ref{fig:3-1-a} with Figure~\ref{fig:3-1-b}. One of them is that, in both cases, the optimized laser power for different $R$ has the constant value: \SI{100}{W}. This result shows that the SNR can be optimized at every mirror radius by applying the same method as at $R =$ \SI{0.5}{m}. However, we have to be aware that the high laser power can cause other noises such as the thermal noise in the cavity mirror.

The second one is that the optimized mirror reflectivity increases with the increase of mirror radius in Figure~\ref{fig:3-1-a}. On the other hand, its value is almost the same in Figure~\ref{fig:3-1-b}. We can explain the reason in the constant mirror-thickness case as follows. In this case, the radiation pressure noise is high for low mirror mass: at the small mirror radius, $R <$ \SI{0.5}{m}. There, as shown in Figure~\ref{fig:3-1-a}, the cavity length has longer values than that of the constant mirror-mass case in order to reduce the radiation pressure noise at the small mirror radius. Nevertheless, this also reduces the value of diffraction $D_\mathrm{opt}$; it has a small value at the high diffraction loss. Consequently, the low mirror reflectivity compensates for the low $D_\mathrm{opt}$ by capturing the laser light inside the FP cavity.

\begin{figure}[H]
   \subfigure[]{
     \includegraphics[clip,width=7.5cm]{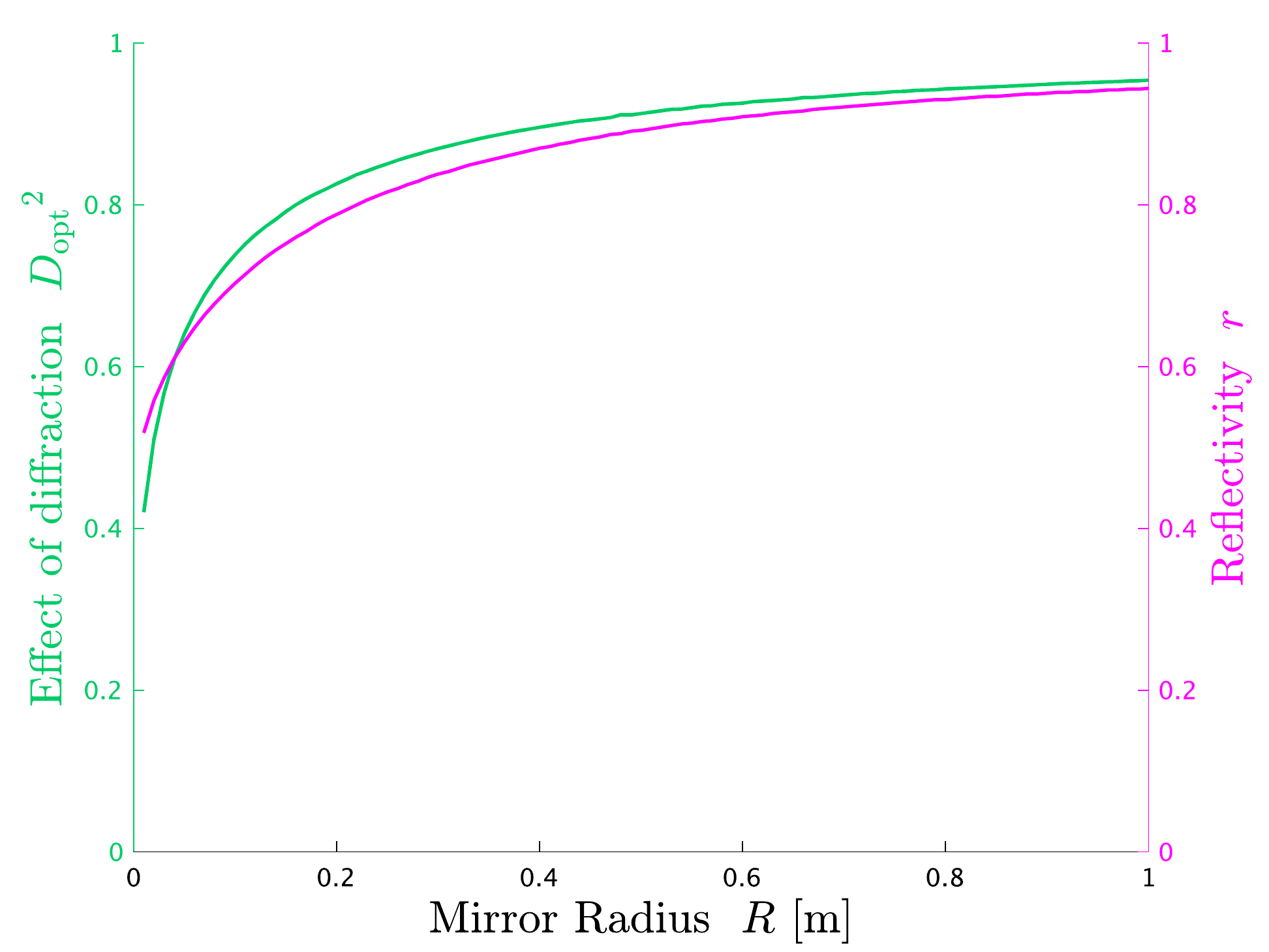}
     \label{fig:3-3-a}
   }
   \hfill
   \subfigure[]{
     \includegraphics[clip,width=7.5cm]{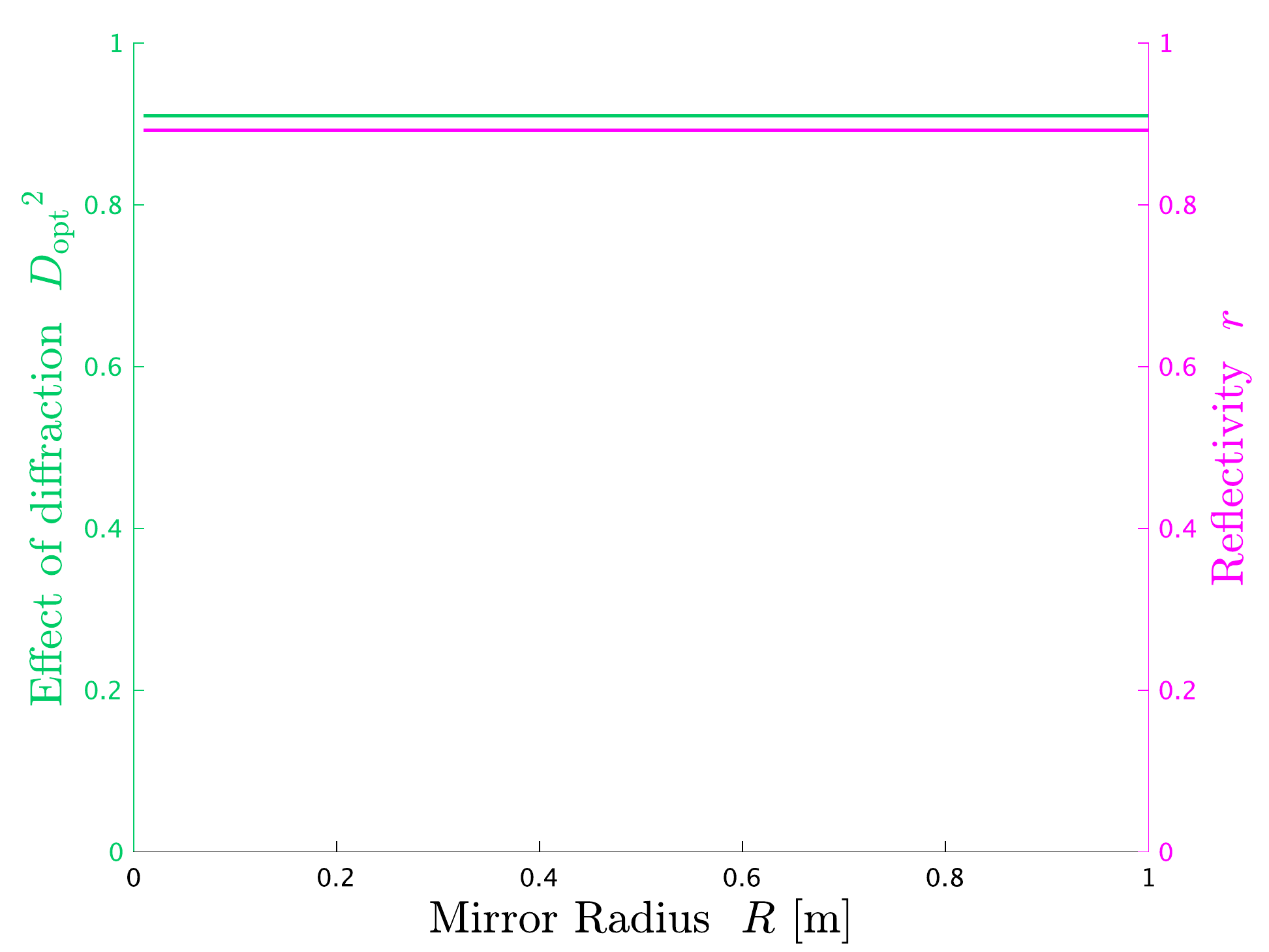}
     \label{fig:3-3-b}
   }
   \vspace{-0.5cm}
   \caption{Optimized effect of diffraction ${D_\mathrm{opt}}^2$ (green line) and reflectivity $r$ (magenta line) for given $R$ in the case of constant thickness (a) and in the case of constant mass (b). Both figures show plotted ${D_\mathrm{opt}}^2$ is similar to that of $r$.}
   \label{fig:3-3}
 \end{figure}

The optimized mirror reflectivity changes for given $R$ in the constant mirror-thickness case in the previous mechanisms. Meanwhile, that of the constant mirror-mass case is almost constant. This is because the optimized $D_\mathrm{opt}$ is almost constant too. Figure~\ref{fig:3-3} shows the optimized effect of diffraction $D_\mathrm{opt}$ as the terms of ${D_\mathrm{opt}}^2$ for different $R$ in bothe cases. In Figure~\ref{fig:3-3-b}, the mirror radius and the cavity length are adjusted in the constant mirror-mass case, and $D_\mathrm{opt}$ has almost identical value at every radius when the parameters are optimized. Consequently, the mirror reflectivity needs not to compensate for the loss.

There is one additional point to note. Compared with the constant mirror-mass case, the optimized SNR at large radius $R >$ \SI{0.5}{m} is much larger in the constant mirror-thickness case. This is caused by a similar process mentioned previously. The heavy mass reduces the radiation pressure noise at the large mirror radius. Then the cavity length is extended, and the shot noise is lowered. Also, the mirror reflectivity has a high value, that is, the finesse is high. Hence, the SNR can have high value in the case of changing mass.

\section{Conclusion}
We have obtained an appropriate combination of DECIGO parameters with diffraction loss by optimizing the SNR of two correlated clusters. In addition, we can enhance the total SNR including the effects of diffraction loss, which we had not considered before. Furthermore, we have discovered a new result that the SNR is enhanced by the cavity wasting a part of the increased laser power. It enables to enhance the total SNR. The target sensitivity is slightly improved by optimizing parameters only. In any cases, the result we obtained here is the first step toward optimizing the DECIGO design by considering the practical constraints on the mirror dimension and implementing other noise sources.

\section*{Acknowledgments}
We would like to thank Naoki Seto for helpful discussion. We also would like to thank Stanley E. Whitcomb for English editing. This work was supported by the Japan Society for the Promotion of Science (JSPS) KAKENHI Grant Number JP19H01924.


\end{document}